# On enhanced hydrogen adsorption on alkali (Cesium) doped $C_{60}$ and effects of the quantum nature of the $H_2$ molecule on physisorption energies

Alexander Kaiser[1*], Michael Renzler[1], Lorenz Kranabetter[1], Matthias Schwärzler[1], Rajendra Parajuli[3], Olof Echt[1,2], Paul Scheier[1]

[1] Institut für Ionenphysik und Angewandte Physik, University of Innsbruck, Technikerstraße 25, A-6020 Innsbruck

[2] Department of Physics, University of New Hampshire, Durham, New Hampshire 03824, USA

[3] Department of Physics, Amrit Campus, Tribhuvan University, Kathmandu, Nepal

*Corresponding author: alexander.kaiser@uibk.ac.at

## Abstract

Hydrogen storage by physisorption in carbon based materials is hindered by low adsorption energies. In the last decade doping of carbon materials with alkali, earth alkali or other metal atoms was proposed as a means to enhance adsorption energies, and some experiments have shown promising results. We investigate the upper bounds of hydrogen storage capacities of $C_{60}$Cs clusters grown in ultracold helium nanodroplets by analyzing anomalies in the ion abundance that indicate shell closure of hydrogen adsorption shells. On bare $C_{60}^+$, a commensurate phase with 32 $H_2$ molecules was identified in previous experiments. Doping $C_{60}$ with a single cesium atom leads to an increase in relative ion abundance for the first 10 $H_2$ molecules, and the closure of the commensurate phase is shifted from 32 to 42 $H_2$ molecules. Density functional theory calculations indicate that thirteen energetically enhanced adsorption sites exist, where six of them fill the groove between Cs and $C_{60}$ and 7 are located at the cesium atom. We emphasize the large effect of the quantum nature of the hydrogen molecule on the adsorption energies, i.e. the adsorption energies are decreased by around 50% for $(H_2)C_{60}$Cs and up to 80% for $(H_2)C_{60}$ by harmonic zero-point corrections, which represent an upper bound to corrections for dissociation energies ($D_e$ to $D_0$) by the vibrational ground states. Five normal modes of libration and vibration of $H_2$ physisorbed on the substrate contribute primarily to this large decrease in adsorption energies. A similar effect can be found for $H_2$ physisorbed on benzene and is expected to be found for any other weakly $H_2$-binding substrate.

## Introduction

Adsorption on nanotubes, fullerenes and other porous carbonaceous materials shows promise for high-density storage of $H_2$ and other hydrogen-rich molecules [1-9] and may be an alternative to other hydrogen-storage technologies [10-16]. However, potential well depths for physisorption are shallow (about 50 meV per $H_2$ molecule for adsorption on bare $C_{60}$) while chemisorption energies are high (about 2.1 eV per H atom for fulleranes, $H_nC_{60}$) [17]; these values are not compatible with the goal of high storage capacity and efficient, reversible hydrogen release at ordinary temperatures. Doping with alkali or earth alkaline metals promises to elevate adsorption energies and storage capacities to 5 wt % or more [18-21].

Doping also stabilizes the substrates against destruction during the ad/desorption process [22, 23]. For alkalis the main mechanism for the increase in physisorption energy is charge transfer from the adsorbate to the degenerate lowest unoccupied molecular orbital of $C_{60}$.

In 2006 Sun et al., using DFT, predicted an astounding hydrogen storage capacity of lithium doped $C_{60}$ [24]. The icosahedral $C_{60}Li_{12}$ was predicted to physisorb 5 $H_2$ molecules per Li which would be equivalent to a hydrogen storage capacity of 13.1 wt %. The hydrogen molecules are bound to $C_{60}(Li)_{12}$ with an average binding energy ($D_e$) of 0.075 eV per $H_2$. A bare neutral lithium atom only binds a single H atom with 2.49 eV, whereas a Li cation binds 6 $H_2$ molecules. A binding energy of 0.202 eV was reported for $H_2$-$Li^+$. Two years later in 2008 Chandrakumar and Gosh calculated $H_2$ adsorption on $C_{60}$ doped with Li, Na or K atoms [20]. They reported that up to six $H_2$ molecules bind to each adsorbed Na; the adsorption energies are enhanced relative to $H_2$ adsorption on bare $C_{60}$ due to the polarization interaction with the adsorbed alkali cations. For comparison, eight $H_2$ can be adsorbed with approximately equal adsorption energies on a free $Na^+$ ion; the presence of $C_{60}$ blocks two of those eight sites. Since then many theoretical studies have been published on hydrogen adsorption on carbonaceous materials doped with alkali, earth alkali, and other metals [25-29].

A more complex behaviour has been reported for adsorption of atomic hydrogen at $C_{60}Li_6$ by Wang and Jena [18]. The lowest energy configuration consists of 5 exohedral Li and one endohedral Li; each exohedral Li physisorbs one $H_2$ molecule. The remaining hydrogen is chemisorbed at the 30 carbon atoms that are not linked to the exohedral Li. Such a complex, $H_{40}C_{60}Li_6$, would have a gravimetric hydrogen storage density of 5 wt %. Experiments performed on bulk samples of lithium- and sodium intercalated $C_{60}$ have proven enhanced values for hydrogen uptake compared to pure $C_{60}$. As much as 5 wt % of $H_2$ could be reversibly chemisorbed and desorbed in $C_{60}Li_6$ and $C_{60}Li_{12}$; the storage capacity of $C_{60}Na_6$ and $C_{60}Na_{10}$ was found to be slightly lower [22, 30-33]. Moreover, the temperature of the onset of desorption was reduced to about 250 - 300 °C, well below the temperature (> 500 °C) at which hydrogen starts to desorb from pure fulleranes [33, 34]. Optical spectroscopy, X-ray diffraction, neutron diffraction, NMR and Muon Spin Relaxation have been applied to probe the dynamics and structural changes that accompany ad- and desorption of hydrogen [23, 30-33, 35-39]. Laser-desorption mass spectra of hydrogenated alkali-doped $C_{60}$ solids show evidence for the presence of $H_nC_{60}$ ($n$ up to 48) fulleranes [23]; other spectra show $H_nC_{60}^+$ ions containing as many as 60 hydrogen atoms [32]. Curiously, though, no trace of $H_nC_{60}$-alkali ions was reported [23, 32].

The present work was stimulated by those reports but here we take an alternative, bottom-up approach. Helium nanodroplets are formed in a supersonic jet and successively doped with $C_{60}$, cesium and $H_2$. Mass spectra of these complexes reveal ions that are particularly stable. For practical purposes we used cesium rather than lithium or sodium because the large mass of cesium avoids any ambiguity when analyzing the composition of the ions.

In particular, we study hydrogen adsorption on $C_{60}$-cesium complexes experimentally and theoretically to address the following questions: i) Does cesium enhance hydrogen uptake in the same way that lighter alkalis do? ii) How many hydrogen atoms exhibit enhanced binding to $C_{60}Cs^+$? iii) What is the effect of a single adsorbed cesium atom on the number of hydrogen atoms in a complete solvation shell surrounding $C_{60}^+$? The experimental data suggest that up to 10 $H_2$ molecules experience enhanced adsorption on $C_{60}Cs^+$ compared to pristine $C_{60}^+$.

Accompanying density functional calculations including dispersion interactions elucidate structure and energetics of the clusters. Adsorption energies are strongly affected by the quantum nature of the light nuclei of $H_2$. Energies decrease significantly upon inclusion of harmonic corrections for the vibrational ground state energies in the complex and its constituents in all degrees of freedom. An estimate for anharmonic effects is given for the radial mode; these also play a notable role for hydrogen adsorption energies. However, more accurate predictions of the vibrational spectrum on a highly accurate multidimensional (at least 5 dimensions) potential energy surface of $(H_2)C_{60}^{0,+}$ would be needed. A Rigid Body Diffusion Quantum Monte Carlo calculation on a fitted model potential revealed a similarly large effect of vibrational zero-point corrections for $H_2$ physisorbed on benzene [40]. The comparison with the benzene model system shows that dispersion corrected density functional calculations with ωB97X-D including harmonic zero-point corrections and the moderate basis set LANL2DZ can provide a qualitative picture for $H_2$ adsorption energies on $C_{60}$, with slightly overestimated zero-point energies (ZPE).

## Experimental results

Sections of a mass spectrum of helium nanodroplets doped with $C_{60}$, Cs, and $H_2$ are displayed in Fig. 1. $H_nC_{60}^+$ ions ($n$ = integer) appear in the upper panel; these ions have been studied before [17, 41]. Cesiumnated $H_nC_{60}Cs^+$ ions are seen in the lower panel. Some mass peaks are labeled. However, the presence of isotopologues needs to be taken into account when interpreting the spectrum. For example, the $H_{20}C_{60}Cs^+$ mass peak contains significant contributions from $H_{19}{}^{13}C^{12}C_{59}Cs^+$, $H_{18}{}^{13}C_2{}^{12}C_{58}Cs^+$ etc., plus a water impurity that is not fully resolved.

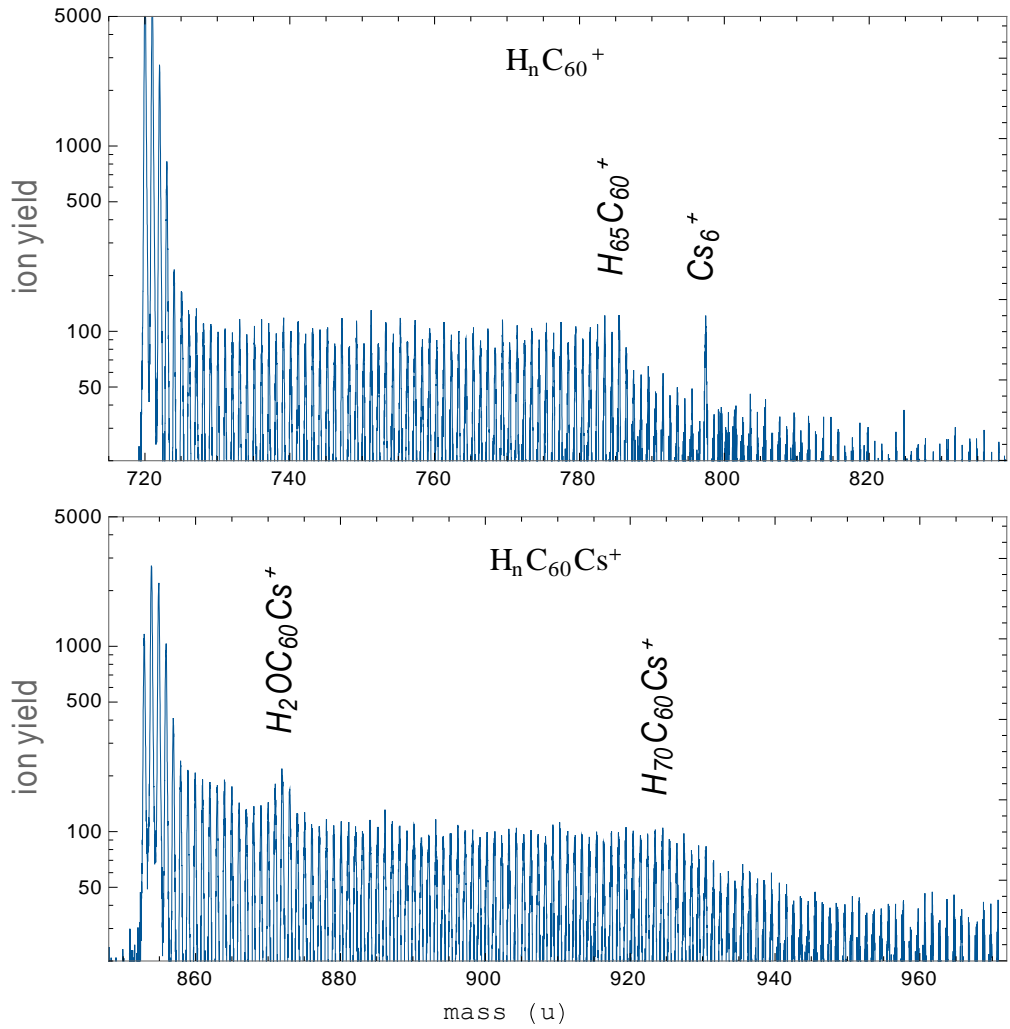

Fig. 1: Two sections of a mass spectrum of helium nanodroplets doped with $C_{60}$, Cs and $H_2$. The upper panel shows the presence of $H_nC_{60}^+$ ions; $H_nC_{60}Cs^+$ ions appear in the lower panel. The water impurity $H_2OC_{60}Cs^+$ and the peak for pure $Cs_6^+$ are labelled as well.

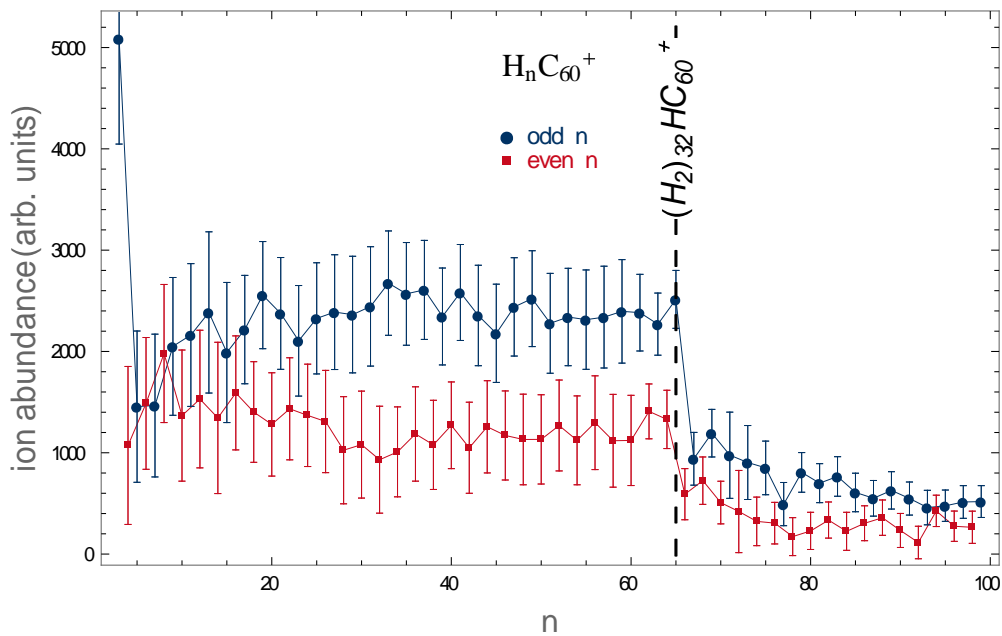

Fig. 2: Abundance distributions of $H_nC_{60}^+$ ions extracted from the spectrum in Fig. 1.

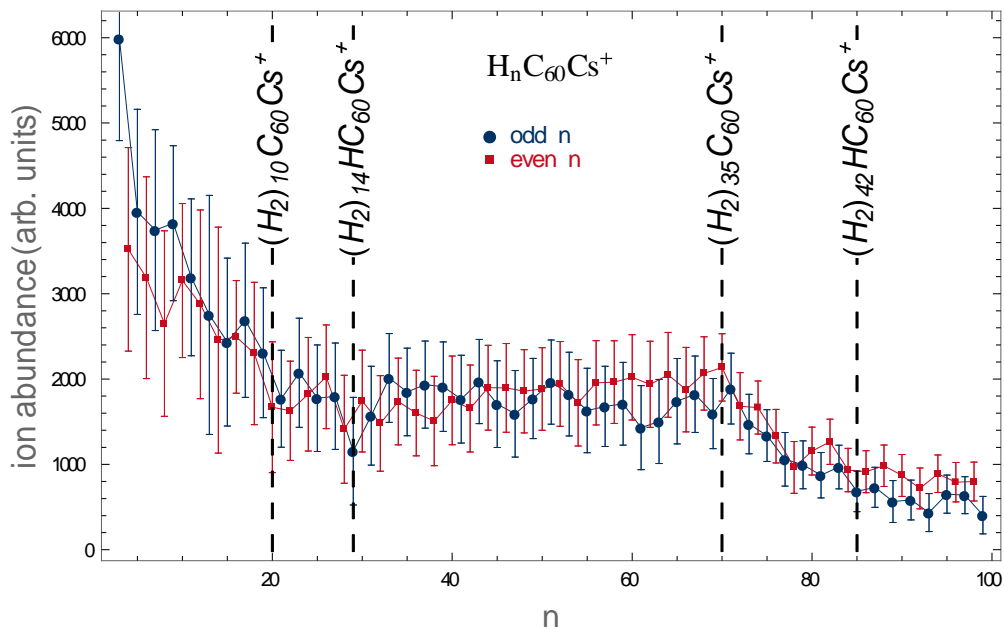

Fig. 3: Abundance distributions of $H_nC_{60}Cs^+$ ions extracted from the spectrum in Fig. 1. Larger error-bars arise from the fitting procedure in case of overlapping series.

Fig. 2 and 3 display the abundance of ions extracted from the mass spectrum (Fig. 1), corrected for the occurrence of isotopologues and impurities using the isotope fit procedure [42]. In the $H_nC_{60}^+$ series, odd-numbered ions are two to three times more abundant than even-numbered ions. This effect has been attributed to the large exothermicity [43] of the gas-phase reaction

$$H_2^+ + H_2 \rightarrow H_3^+ + H + 1.727 \text{ eV}. \qquad (2)$$

Although the energetics will be different in the presence of $C_{60}$ and other hydrogen molecules, ionization of a fullerene coated with a layer of molecular hydrogen will in all likelihood produce $H_3^+$. Density functional calculations showed that the proton affinity of $C_{60}$ (9.27 eV) is large enough to cause a barrierless reaction with $H_3^+$ to an equilibrium geometry where one H binds covalently to $C_{60}$, now positively charged, and $H_2$ is physisorbed [17].

In contrast to $H_nC_{60}^+$ and pure $H_n^+$ clusters, there is almost no odd/even difference in the ion abundance of $H_nC_{60}Cs^+$. This is likely a consequence of the ionization processes. In the case of $H_nC_{60}^+$ ionization primarily proceeds inside the droplet via $He^+$ formation and resonant hole hopping or via highly mobile $He^{*-}$ [44]. In the case of $H_nC_{60}Cs^+$, we first doped the helium droplets with fullerene and hydrogen which can only form neutral $(H_2)_mC_{60}$ clusters with a purely statistical distribution and no odd numbered species. Cesium was added in a second pickup chamber and, since cesium is a heliophobic species [45-48], it resides preferentially on the surface of the droplet [49]. Thus, another ionization process is accessible, where cesium is ionized on the outside of the droplet via Penning ionization by He* [50], which is also a heliophobic species [51]. $Cs^+$ reacts with neutral $(H_2)_mC_{60}$ by polarization induced interaction. This leads only to even numbered $H_nC_{60}Cs^+$ clusters, since the reaction $Cs^+ + H_2 \Rightarrow CsH^+ + H$ is highly endothermic (ca. 4.4 eV according to our calculations). On the other hand, if $H_2$ is ionized via $He^+$ or $He^{*-}$, $CsH_2^+$ and $CsH^+ + H$ can be formed with exothermicities of 11.7 eV and 7.3 eV respectively. These mechanisms compete with direct ionization of $C_{60}$ and with reaction (2). In summary, a variety of possible ionization channels exist, where Penning ionization leads to exclusively even numbered clusters and seems to balance the scale between odd and even numbered clusters by chance.

Besides differences in the odd/even behavior, remarkable differences caused by cesium also arise for small *n* and near the second drop in the abundance in Figs. 2 and 3. The abrupt drop in the ion abundance at $H_{65}C_{60}^+$ (or at $H_{64}C_{60}^+$ for the even-numbered series, see Fig. 2) occurs when all 32 hexagonal and pentagonal faces of $C_{60}$ are decorated by one $H_2$ each. A more gradual drop occurs between *n* = 70 and *n* = 85 in the abundance of *cesiumnated* $H_nC_{60}Cs^+$ in Fig. 3. We also observe a much higher and gradually decreasing abundance for 5 ≤ *n* ≤ 20 for the cesiumnated species compared to the almost constant (even *n*) or slightly increasing abundance (odd *n*) for species without cesium. A minimum in the abundance at *n* = 29 is probably not significant within the error bars.

We tentatively interpret these observations as follows: A single cesium ion will most likely adsorb above the center of a hexagonal or pentagonal carbon ring (calculations for the lighter alkalis indicate that sodium and potassium preferentially adsorb at a hexagonal site while for lithium the pentagonal and hexagonal sites are nearly isoenergetic [52, 53]). According to the DFT study by Chandrakhumar and Ghosh, a maximum of 8 $H_2$ molecules bind strongly to an isolated $Na^+$ or $K^+$ ion, while a Na atom bound to $C_{60}$ offers 6 adsorption sites for $H_2$ [20]. A slightly increased binding energy is also expected for adsorption sites on $C_{60}$ near the Cs ion. In total we count 6 to 8 adsorption sites localized on $Cs^+$, 6 adsorption sites on $C_{60}$ near $Cs^+$, and 25 regular, hollow sites on $C_{60}$ with binding energies close to those on undoped $C_{60}^+$. This theoretically gives 12-14 enhanced adsorption sites (for 24-28 H atoms) and 25 regular sites (50 H) on $C_{60}$. We also have to consider that for a given coverage number *n* below full coverage, a huge number of isomers with slightly different binding energies can be formed according to the different binding sites and therefore we cannot expect a sharp drop in the ion abundance as observed for undoped $C_{60}$ (see Fig. 2) for the commensurate (*n* = 2×32) phase. The gradual decrease for 5 ≤ *n* ≤ 20 for the cesiumnated complex points to around 10 stronger adsorption sites near $Cs^+$ on $C_{60}$. These 10 sites plus 25 hollow sites over centers of hexagons and pentagons lead to a commensurate coverage for $(H_2)_{35}C_{60}Cs^+$, which explains the onset of the second gradual decrease in the ion abundance in Fig. 3. However, also for 70 < *n* ≤ 84, the abundance is much higher than in the non-cesiumnated system, indicating that ca. 7 more $H_2$ can be squeezed in near $Cs^+$ with gradually decreasing (but still enhanced) binding energy. In summary, we observe a higher ion yield caused by cesium for 5 ≤ *n* ≤ 20 and 65 ≤ *n* ≤ 84 hydrogen atoms which can be explained in terms of up to 10 additional $H_2$ adsorption sites compared to bare $C_{60}$.

## Theoretical Results

We have performed density functional theory calculations including empirical dispersion correction (see section Experimental and Theoretical Details) in order to shed light on the possible structures and energetics of cationic $(H_2)_mC_{60}Cs^+$ clusters. Similar theoretical studies have been published before for lighter alkali atoms such as Li, Na, K [20, 24, 29]. Qualitatively, we expect small differences to previous calculations caused by the large size of cesium. A positive net charge of +1 e is expected to be localized on Cs due to its low ionization energy, whereas the fullerene remains neutral for cationic clusters. In a standard Mulliken charge analysis we find a net charge of +0.999 e on Cs for $C_{60}Cs^+$, and +0.975 e on Cs for neutral $C_{60}Cs$. A summary of endothermic reactions and their reaction energies at 0 K is given in Table 1, where ΔE is the difference in the total energies of reactants and products and $ΔE^{ZPE}$ also includes harmonic vibrational zero-point corrections. Basis set superposition errors were not corrected for in this study.

Table 1. Calculated endothermic reaction energies at 0 K without (ΔE) and with ($ΔE^{ZPE}$) harmonic zero-point vibrational energy corrections without frequency scaling at the ωB97X-D/LANL2DZ level of theory.

| Reaction | ΔE (eV) | $ΔE^{ZPE}$ (eV) |
| --- | --- | --- |
| CsH => Cs + H | 1.393 | 1.340 |
| $CsH^+$ => $Cs^+$ + H | 0.049 | 0.041 |
| $CsH^+$ + H => Cs + $H_2^+$ | 7.208 | 7.317 |
| $Cs_2$ => Cs + Cs | 0.734 | 0.730 |
| $Cs_2^+$ => $Cs^+$ + Cs | 0.679 | 0.677 |
| $C_{60}Cs$ => $C_{60}$ + Cs | 1.966 | 2.086 |
| $C_{60}Cs^+$ => $C_{60}$ + $Cs^+$ | 0.795 | 0.793 |
| $C_{60}Cs^+$ => $C_{60}^+$ + Cs | 5.406 | 5.300 |
| $CsH_2^+$ => $Cs^+$ + $H_2$ | 0.101 | 0.056 |
| $CsH_2^+$ => Cs + $H_2^+$ | 11.958 | 11.753 |
| $Cs^+$ + $H_2$ => $CsH^+$ + H | 4.648 | 4.380 |
| Cs + $H_2$ => CsH + H | 3.304 | 3.081 |
| $Cs_2$ + $H_2$ => 2 CsH | 2.644 | 2.470 |
| $KH_2^+$ => $K^+$ + $H_2$ | 0.061 | 0.041 |
| $(H_2)C_6H_6$ => $C_6H_6$ + $H_2$ | 0.045 | negative (-0.0001) |
| $(H_2)C_{60}$ => $C_{60}$ + $H_2$ | 0.051 | 0.010 |
| $(H_2)C_{60}^+$ => $C_{60}^+$ + $H_2$ | 0.060 | 0.013 |
| $(H_2)C_{60}Cs$ => $C_{60}Cs$ + $H_2$ | 0.119 | 0.060 |
| $(H_2)C_{60}Cs^+$ => $C_{60}Cs^+$ + $H_2$ | 0.110 | 0.052 |

The Coulomb attraction between Cs and $C_{60}$ leads to a strong interaction of 2.09 eV in the neutral case, whereas $C_{60}Cs^+$ is bound by 0.8 eV only. For comparison, somewhat lower binding energies of 1.71 eV for $C_{60}Cs$ and 0.48 eV for $C_{60}Cs^+$ are obtained with the PBE0 hybrid functional and Grimme D3 dispersion. Similar to other alkali species [54], $H_2$ does not bind well with neutral Cs, whereas for $CsH_2^+$ our DFT calculations predict a positive potential well depth $D_e$ = 0.101 eV, which is remarkably diminished to $D_0$ = 0.056 eV including harmonic ZPE. Larger binding energies of 0.253 eV ($D_e$) for $LiH_2^+$ with MP4/6-311G(d,p) [54] and 0.165 eV for $NaH_2^+$ with B3LYP/6-31+g(d,p) [20] have been reported for lighter alkali atoms. Bodrenko et al. calculated $H_2$ binding energies to benzene-alkali complexes with MP2/AVTZ and obtained 0.156 eV for $C_6H_6LiH_2$, 0.083 eV for $C_6H_6NaH_2$, and 0.052 eV for $C_6H_6KH_2$ without ZPE [27]; these values decrease significantly from light Li to heavier K. With the method employed in this work, we obtain 0.061 eV (0.041 including ZPE) for pristine $KH_2^+$ which is in good agreement with the value reported by Bodrenko et al. for the benzene-$KH_2$ complex.

On pristine neutral and charged $C_{60}$ we obtain $H_2$ binding energies ($D_e$) of 0.051 eV and 0.060 eV respectively, when $H_2$ is adsorbed flat over centers of hexagonal faces. These values are very similar to the values of 0.050 and 0.057 eV respectively that we reported in Ref. [55] with a different basis set. In that work we estimated a ZPE correction of 0.011 eV for the radial normal mode only [55]. By fully including harmonic zero-point vibrational energy corrections for all degrees of freedom, we obtain a significant decrease of the adsorption energy to $D_0$ = 0.010 eV for $(H_2)C_{60}$ and $D_0$ = 0.013 eV for $(H_2)C_{60}^+$. This arises from five normal modes (excluding the stretching mode) of librations and vibrations of $H_2$ on $C_{60}$. A similarly large effect of the quantum nature of the hydrogen molecule has been obtained by Diffusion Quantum Monte Carlo calculations for $H_2$ physisorbed on benzene; in this system $D_e$ = 0.040 eV decreased to $D_0$ = 0.014 eV [40]. With the level of theory employed in the present work, we found $D_e$ = 0.045 eV for $H_2$-benzene, but a slightly negative value for $D_0$, which shows the limitations of the present approach.

For high-lying vibrational ground states the anharmonicity of the potential energy surface can be expected to play a significant role. From the data of Ref. [55] we can estimate an anharmonic contribution of 0.003 eV for the radial normal mode only. Assuming similar magnitudes for the other vibrational modes would also give a non-negligible contribution to $D_0$. Scanlon et al. have calculated, with MP2(Full)/6-311++G(3df,2p), enthalpies of adsorption for the $H_2$-benzene system at 298 K and reported ΔH = 0.011 eV given a binding energy ($D_e$) of 0.052 eV [56].

The relatively large ZPE correction by including all degrees of freedom of $H_2$ nuclei could indeed lead to adsorption energies ($D_0$ or ΔH) for $H_2$ on carbonaceous structures that are similar to adsorption energies for He, the latter being heavier and having fewer degrees of freedom for librations and vibrations [57]. Further investigation is needed for a reliable quantitative description. Here, we are more interested in the structures and in qualitative trends in the energetics coming from the polarization induced interaction with the cesium ion and do not intend to report highly accurate $C_{60}H_2$ interaction energies which would, in any case, demand higher accuracy for electron correlation and a larger basis sets.

Dissociation energies are given in Table 2 for $H_2$ desorption; they are defined by $D_e(m) = E(H_2) + E((H_2)_{m-1}C_{60}Cs^+) - E((H_2)_m C_{60}Cs^+)$, where $E$ denotes total energies of locally optimized structures. Harmonic vibrational zero-point corrections are included in $D_0$. $H_2$ prefers to adsorb on $C_{60}Cs^+$ in such a way that it maximizes its polarization interaction with $Cs^+$ and its van der Waals interaction with $C_{60}$ (see Figure 4) which results in $D_e$ = 0.110 eV ($D_0$ = 0.061 eV). We call this a groove site (Figure 4c) in analogy to the grooves between two or more fullerenes/nanotubes [58, 59]. Marginally smaller values of $D_e$ = 0.080 eV are reached outside the groove near Cs. Although the energetic difference is hardly significant within the accuracy of this level of theory, this trend seems reasonable. In total, the groove prefers to accommodate up to 6 $H_2$ molecules (Figure 4a and 4b) with $D_e$ values between 0.110 and 0.118 eV, which is slightly larger than the calculated $CsH_2^+$ well depth of $D_e$ = 0.101 eV. Here the $H_2$-$H_2$ interaction also contributes in addition to $H_2C_{60}$ van der Waals forces. The closure of the groove site at $m$ = 6 gives a particularly symmetric structure, where every $H_2$ resides over a hexagonal or pentagonal face adjacent to $Cs^+$ in an alternating orientation (Figure 4a and 4b). Adding more than 6 $H_2$ leads to decoration of the cesium atom away from the groove site with binding energies between 0.091 eV ($m$ = 7) and 0.100 eV ($m$ = 12, Figure 4d). The dissociation energy decreases again for $m ≥ 13$, which we associate with the closure of the first $H_2$ adsorption shell around cesium. This agrees well with the very simple considerations described above, which indicated $m$ = 12 to 14 stronger adsorption sites near Cs, giving, including 25 hollow sites on $C_{60}$, 37 to 39 sites in total. It also compares well within the error bars of the experimental data with the observation of up to 42 adsorption sites. In the simulations we found that the 15[th] $H_2$ adsorbed on $C_{60}Cs^+$ was the first one not bound near Cs during local optimization.

Table 2. Calculated dissociation energies for $(H_2)_m C_{60}Cs^+$ without ($D_e$) and with harmonic corrections for zero-point vibrations ($D_0$) without frequency scaling at the ωB97X-D/LANL2DZ level of theory. For comparison, $D_e$ = 0.096 eV and $D_0$ = 0.044 eV are obtained for m = 1 with the PBE0 functional and Grimme D3 correction.

| m | $D_e$ [eV] | $D_0$ [eV] | m | $D_e$ [eV] | $D_0$ [eV] |
|---|---|---|---|---|---|
| 1 | 0.110 | 0.061 | 8 | 0.091 | 0.042 |
| 2 | 0.115 | 0.075 | 9 | 0.096 | 0.058 |
| 3 | 0.116 | 0.055 | 10 | 0.099* | - |
| 4 | 0.115 | 0.062 | 11 | 0.090 | - |
| 5 | 0.117 | 0.063 | 12 | 0.100 | 0.046 |
| 6 | 0.118 | 0.074 | 13 | 0.083 | 0.016 |
| 7 | 0.091 | 0.049 | 14 | 0.054 | not bound |

* For m = 10 we could only find a first order saddle point by optimization, all other reported structures are local minima

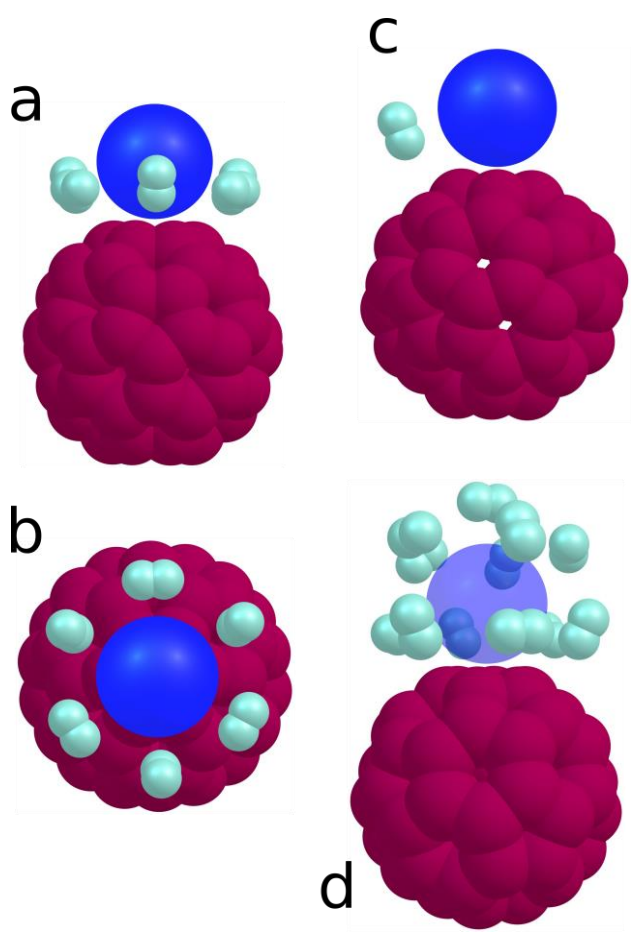

Figure 4. Front (a) and top (b) view of $(H_2)_6C_{60}Cs^+$, and front views of optimized $H_2C_{60}Cs^+$ (c) and $(H_2)_{12}C_{60}Cs^+$ (d).

## Conclusion

For cesiumnated $C_{60}$ cations we could unambiguously identify an increase in the relative ion abundance for $m \leq 10$ and for $32 \leq m \leq 42$ hydrogen molecules compared to bare $C_{60}^+$ at the very cold temperatures in helium nanodroplets. This confirms that cesium ions provide stronger $H_2$ adsorption than $C_{60}^+$. Unlike a sharp drop in the ion abundance for the pristine $(H_2)_nC_{60}^+$ system, we see a gradual decrease for the cesiumnated system which can be attributed to a large number of possible metastable isomers with different stability. Taking the conservative estimate of 39 adsorption sites of $H_2$ on $C_{60}Cs^+$ from our theoretical considerations which is in line with the experiment within its error bars, each Cs atom will add another 7 adsorption sites, assuming that Cs atoms adsorb well separated from each other and each Cs transfers its valence electron to the LUMO of $C_{60}^+$ [20, 49]. This gives rise to adsorption of 39, 46, 53, 60, 67 $H_2$ for 1, 2, 3, 4, 5 cesium respectively, corresponding to 8.4, 8.6, 8.7, 8.8, 8.9 wt % gravimetric storage capacity. These values are only slightly larger than for the commensurate phase on bare $C_{60}^+$ (8.2 wt %), but the adsorption energies are larger. The situation is different for lighter alkalis for which the gravimetric storage capacity can increase significantly, up to 13 wt % for $C_{60}Li_{12}$ [24]. Although the binding energies are very similar, Li is much lighter than Cs. From this point of view lithium doping is a much better choice for $H_2$ storage applications than cesium doping.

We have elaborated on the importance to include the quantum nature of $H_2$ molecules in calculating adsorption energies for weakly binding substrates. In particular, $D_0$ and $\Delta H$ can be significantly smaller than the potential well depth $D_e$. In particular, $D_0$ is weaker than $D_e$ by ca. 80 % for $(H_2)C_{60}$ and $(H_2)C_{60}^+$, and by 44 % to 53 % for $(H_2)C_{60}Cs$, $(H_2)C_{60}Cs^+$, $CsH_2^+$ by inclusion of harmonic zero point corrections. Path integral molecular dynamics with a rotationally averaged potential has been used recently by Calvo and Yurtsever to calculate the influence of nuclear quantum effects on the closure of the first $H_2/D_2$ solvation shell of $C_{60}$. A similar description for the cesiumnated system would be highly appreciated for a more rigorous treatment of these effects [60].

# Experimental and theoretical details

Neutral helium nanodroplets were produced by expanding helium (Messer, purity 99.9999 %) at a stagnation pressure of 20 bar through a 5 μm nozzle, cooled by a closed-cycle refrigerator to 9.4 K, into a vacuum chamber (base pressure about $4\times10^{-6}$ Pa). Droplets that form in the expansion contain an average number of $5\times10^5$ helium atoms [61]; the droplets are superfluid with a temperature of 0.37 K [62]. The resulting supersonic beam was skimmed by a 0.8 mm conical skimmer, located 8 mm downstream from the nozzle. The skimmed beam traversed a 20 cm long pick-up region consisting of two separate pick-up chambers [41]. Small amounts of $C_{60}$ (MER, purity 99.9 %) were vaporized in the first chamber. The second chamber contained cesium vapor produced by vaporization of metallic cesium (Sigma Aldrich, purity 99.95 % based on a trace metals analysis) at about 60 °C, and hydrogen gas. The partial pressures of $C_{60}$, cesium and $H_2$ were varied to obtain the desired ratio of these species in the mass spectra.

The beam emerging from the dual pickup cell was collimated and crossed by an electron beam with an energy of 26 eV. Cations were accelerated into the extraction region of a reflectron time-of-flight mass spectrometer (Tofwerk AG, model HTOF) with a mass resolution $\Delta m/m$ = 1/5000 ($\Delta m$ = full-width-at-half-maximum). The base pressure in the mass spectrometer was $10^{-5}$ Pa. Ions were extracted at 90° into the field-free region of the spectrometer by a pulsed voltage. At the end of the field-free region they entered a two-stage reflectron which reflected them towards a microchannel plate detector operated in single ion counting mode. Additional experimental details have been provided elsewhere [63].

Mass spectra were evaluated by means of a custom-designed software [17, 64]. The routine includes automatic fitting of Gaussians to the mass peaks and subtraction of background by fitting a spline to the background level of the raw data. It explicitly considers the presence of $He_n^+$, a variety of impurity ions (e.g., OHC$_{60}^+$) and isotopic patterns. Cesium is monisotopic and hydrogen is very nearly so (the natural abundance of deuterium is 0.0115 %). However, the natural abundance of $^{13}$C is 1.07 %; hence 48 % of all C$_{60}$ ions have a mass exceeding 720 u. Taking these isotopologues into account is crucial when analyzing complexes of C$_{60}$ with hydrogen. The abundance of ions (i.e. ions of a specific composition, e.g. H$_{65}$C$_{60}$Cs$^+$) is derived by a matrix method.

The ωB97X-D hybrid density functional [65] empirically accounts for long-ranged dispersion using Grimme's method [66] and was already used in our group for estimating energetics of hydrogen adsorption on pristine and protonated fullerenes [55]. Energetics of selected structures were compared to values obtained with the PBE0 hybrid functional including Grimme D3 correction [67-70]. In the present work, we employed the double zeta basis set LANL2DZ including an effective core potential (ECP) for Cs [71, 72] as given in the EMSL basis set exchange library [73]. All structures were locally optimized and frequency calculations confirmed that we have reached true local minima; frequencies were not scaled for ZPE calculation. Basis set superposition errors were not taken into account here; their magnitude was estimated with 0.012 eV for (H$_2$)C$_{60}$ in Ref. [55]. All calculations were performed with the Gaussian 09 code (version D.01) [74].

# Acknowledgments

This work was supported by the Austrian Science Fund, Wien (FWF Projects I978, P23657, P26635, P28979-N27). The computational results presented have been achieved (in part) using the HPC infrastructure LEO of the University of Innsbruck. A.K. acknowledges support by the Tirolean Science Fund (TWF grant). R.P. thanks the Austrian Agency for International Cooperation in Education and Research ORGANISATION (OeAD-GmbH), Centre for International Cooperation & Mobility (ICM) for Ernst Mach-Nachbetreuungsstipendium (EZA)

# References

[1] N.S. Venkataramanan, H. Mizuseki, Y. Kawazoe, Hydrogen Storage on Nanofullerene Cages, Nano, 4 (2009) 253-263. http://dx.doi.org/10.1142/s1793292009001733
[2] M. Terrones, A.R. Botello-Méndez, J. Campos-Delgado, F. López-Urías, Y.I. Vega-Cantú, F.J. Rodríguez-Macías, A.L. Elías, E. Muñoz-Sandoval, A.G. Cano-Márquez, J.-C. Charlier, H. Terrones, Graphene and graphite nanoribbons: Morphology, properties, synthesis, defects and applications, Nano Today, 5 (2010) 351-372. http://dx.doi.org/doi:10.1016/j.nantod.2010.06.010
[3] P. Jena, Materials for Hydrogen Storage: Past, Present, and Future, J Phys Chem Lett, 2 (2011) 206-211. http://dx.doi.org/10.1021/jz1015372
[4] M. Pumera, Graphene-based nanomaterials for energy storage, Energy Environ Sci, 4 (2011) 668-674. http://dx.doi.org/10.1039/c0ee00295j
[5] N. Park, K. Choi, J. Hwang, D.W. Kim, D.O. Kim, J. Ihm, Progress on first-principles-based materials design for hydrogen storage, Proc Natl Acad Sci U S A, 109 (2012) 19893-19899. http://dx.doi.org/10.1073/pnas.1217137109
[6] Y. Gao, X.J. Wu, X.C. Zeng, Designs of fullerene-based frameworks for hydrogen storage, J Mater Chem A, 2 (2014) 5910-5914. http://dx.doi.org/10.1039/c3ta13426a
[7] S. Niaz, T. Manzoor, A.H. Pandith, Hydrogen storage: Materials, methods and perspectives, Renewable Sustainable Energy Rev, 50 (2015) 457-469. http://dx.doi.org/http://dx.doi.org/10.1016/j.rser.2015.05.011

[8] F. Calvo, E. Yurtsever, Solvation of carbonaceous molecules by para-H2 and ortho-D2 clusters. I. Polycyclic aromatic hydrocarbons, J Chem Phys, 144 (2016) 224302.
http://dx.doi.org/doi:http://dx.doi.org/10.1063/1.4952957
[9] R. Ströbel, J. Garche, P.T. Moseley, L. Jörissen, G. Wolf, Hydrogen storage by carbon materials, Journal of Power Sources, 159 (2006) 781-801. http://dx.doi.org/http://dx.doi.org/10.1016/j.jpowsour.2006.03.047
[10] B. Sakintuna, F. Lamari-Darkrim, M. Hirscher, Metal hydride materials for solid hydrogen storage: A review, Int J Hydrogen Energy, 32 (2007) 1121-1140.
http://dx.doi.org/http://dx.doi.org/10.1016/j.ijhydene.2006.11.022
[11] P. Chen, M. Zhu, Recent progress in hydrogen storage, Materials Today, 11 (2008) 36-43.
http://dx.doi.org/http://dx.doi.org/10.1016/S1369-7021(08)70251-7
[12] I.P. Jain, P. Jain, A. Jain, Novel hydrogen storage materials: A review of lightweight complex hydrides, Journal of Alloys and Compounds, 503 (2010) 303-339.
http://dx.doi.org/http://dx.doi.org/10.1016/j.jallcom.2010.04.250
[13] N.A.A. Rusman, M. Dahari, A review on the current progress of metal hydrides material for solid-state hydrogen storage applications, Int J Hydrogen Energy, 41 (2016) 12108-12126.
http://dx.doi.org/http://dx.doi.org/10.1016/j.ijhydene.2016.05.244
[14] J. Dong, X. Wang, H. Xu, Q. Zhao, J. Li, Hydrogen storage in several microporous zeolites, Int J Hydrogen Energy, 32 (2007) 4998-5004. http://dx.doi.org/http://dx.doi.org/10.1016/j.ijhydene.2007.08.009
[15] L.J. Murray, M. Dinca, J.R. Long, Hydrogen storage in metal-organic frameworks, Chem Soc Rev, 38 (2009) 1294-1314. http://dx.doi.org/10.1039/B802256A
[16] A. Staubitz, A.P.M. Robertson, I. Manners, Ammonia-Borane and Related Compounds as Dihydrogen Sources, Chem Rev, 110 (2010) 4079-4124. http://dx.doi.org/10.1021/cr100088b
[17] A. Kaiser, C. Leidlmair, P. Bartl, S. Zöttl, S. Denifl, A. Mauracher, M. Probst, P. Scheier, O. Echt, Adsorption of Hydrogen on Neutral and Charged Fullerene: Experiment and Theory, J Chem Phys, 138 (2013) 074311.
http://dx.doi.org/DOI: 10.1063/1.4790403
[18] Q. Wang, P. Jena, Density functional theory study of the interaction of hydrogen with $Li_6C_{60}$, J Phys Chem Lett, 3 (2012) 1084-1088. http://dx.doi.org/10.1021/jz3002037
[19] M. Yoon, S.Y. Yang, C. Hicke, E. Wang, D. Geohegan, Z.Y. Zhang, Calcium as the superior coating metal in functionalization of carbon fullerenes for high-capacity hydrogen storage, Phys Rev Lett, 100 (2008) 206806.
http://dx.doi.org/10.1103/PhysRevLett.100.206806
[20] K.R.S. Chandrakumar, S.K. Ghosh, Alkali-metal-induced enhancement of hydrogen adsorption in $C_{60}$ fullerene: An ab initio study, Nano Lett, 8 (2008) 13-19. http://dx.doi.org/Doi 10.1021/Ni071456i
[21] Z. Ozturk, C. Baykasoglu, M. Kirca, Sandwiched graphene-fullerene composite: A novel 3-D nanostructured material for hydrogen storage, Int J Hydrogen Energy, 41 (2016) 6403-6411.
http://dx.doi.org/http://dx.doi.org/10.1016/j.ijhydene.2016.03.042
[22] P. Mauron, A. Remhof, A. Bliersbach, A. Borgschulte, A. Züttel, D. Sheptyakov, M. Gaboardi, M. Choucair, D. Pontiroli, M. Aramini, A. Gorreri, M. Ricco, Reversible hydrogen absorption in sodium intercalated fullerenes, Int J Hydrogen Energy, 37 (2012) 14307-14314. http://dx.doi.org/10.1016/j.ijhydene.2012.07.045
[23] J.A. Teprovich, M.S. Wellons, R. Lascola, S.J. Hwang, P.A. Ward, R.N. Compton, R. Zidan, Synthesis and characterization of a lithium-doped fullerane ($Li_xC_{60}H_y$) for reversible hydrogen storage, Nano Lett, 12 (2012) 582-589. http://dx.doi.org/10.1021/nl203045v
[24] Q. Sun, P. Jena, Q. Wang, M. Marquez, First-principles study of hydrogen storage on $Li_{12}C_{60}$, J Am Chem Soc, 128 (2006) 9741-9745.
[25] C. Tang, F. Gao, Z. Zhang, J. Kang, J. Zou, Y. Xu, W. Zhu, The properties of hydrogenated derivatives of the alkali atom coated clusters $C_6M_6$ (M = Li, Na): A density functional study, Comput Theor Chem, 1071 (2015) 46-52. http://dx.doi.org/http://dx.doi.org/10.1016/j.comptc.2015.06.023
[26] T. Yildirim, S. Ciraci, Titanium-Decorated Carbon Nanotubes as a Potential High-Capacity Hydrogen Storage Medium, Phys Rev Lett, 94 (2005) 175501.
[27] I.V. Bodrenko, A.V. Avdeenkov, D.G. Bessarabov, A.V. Bibikov, A.V. Nikolaev, M.D. Taran, E.V. Tkalya, Hydrogen Storage in Aromatic Carbon Ring Based Molecular Materials Decorated with Alkali or Alkali-Earth Metals, J Phys Chem C, 116 (2012) 25286-25292. http://dx.doi.org/10.1021/jp305324p
[28] W. Liu, Y.H. Zhao, Y. Li, Q. Jiang, E.J. Lavernia, Enhanced Hydrogen Storage on Li-Dispersed Carbon Nanotubes, J Phys Chem C, 113 (2009) 2028-2033. http://dx.doi.org/10.1021/jp8091418
[29] M. Robledo, S. Diaz-Tendero, F. Martin, M. Alcami, Theoretical study of the interaction between molecular hydrogen and [MC60]+ complexes, RSC Advances, 6 (2016) 27447-27451.
http://dx.doi.org/10.1039/C6RA00501B

[30] A. Paolone, F. Vico, F. Teocoli, S. Sanna, O. Palumbo, R. Cantelli, D.A. Knight, J.A. Teprovich, R. Zidan, Relaxation Processes and Structural Changes in Li- and Na-Doped Fulleranes for Hydrogen Storage, J Phys Chem C, 116 (2012) 16365-16370. http://dx.doi.org/10.1021/jp304552r
[31] J.A. Teprovich, D.A. Knight, B. Peters, R. Zidan, Comparative study of reversible hydrogen storage in alkali-doped fulleranes, J Alloys and Compounds, 580 (2013) S364-S367. http://dx.doi.org/10.1016/j.jallcom.2013.02.024
[32] P.A. Ward, J.A. Teprovich, R.N. Compton, V. Schwartz, G.M. Veith, R. Zidan, Evaluation of the physi- and chemisorption of hydrogen in alkali (Na, Li) doped fullerenes, International Journal of Hydrogen Energy, 40 (2015) 2710-2716. http://dx.doi.org/10.1016/j.ijhydene.2014.12.033
[33] P. Mauron, M. Gaboardi, D. Pontiroli, A. Remhof, M. Ricco, A. Züttel, Hydrogen Desorption Kinetics in Metal Intercalated Fullerides, J Phys Chem C, 119 (2015) 1714-1719. http://dx.doi.org/10.1021/jp511102y
[34] A.A. Peera, L.B. Alemany, W.E. Billups, Hydrogen storage in hydrofullerides, Appl Phys A, 78 (2004) 995-1000. http://dx.doi.org/10.1007/s00339-003-2420-1
[35] P. Mauron, M. Gaboardi, A. Remhof, A. Bliersbach, D. Sheptyakov, M. Aramini, G. Vlahopoulou, F. Giglio, D. Pontiroli, M. Ricco, A. Züttel, Hydrogen Sorption in Li12C60, J Phys Chem C, 117 (2013) 22598-22602. http://dx.doi.org/10.1021/jp408652t
[36] F. Giglio, D. Pontiroli, M. Gaboardi, M. Aramini, C. Cavallari, M. Brunelli, P. Galinetto, C. Milanese, M. Ricco, $Li_{12}C_{60}$: A lithium clusters intercalated fulleride, Chem Phys Lett, 609 (2014) 155-160. http://dx.doi.org/10.1016/j.cplett.2014.06.036
[37] M. Gaboardi, S. Duyker, C. Milanese, G. Magnani, V.K. Peterson, D. Pontiroli, N. Sharma, M. Ricco, In Situ Neutron Powder Diffraction of $Li_6C_{60}$ for Hydrogen Storage, J Phys Chem C, 119 (2015) 19715-19721. http://dx.doi.org/10.1021/acs.jpcc.5b06711
[38] L. Maidich, D. Pontiroli, M. Gaboardi, S. Lenti, G. Magnani, G. Riva, P. Carretta, C. Milanese, A. Marini, M. Ricco, S. Sanna, Investigation of Li and H dynamics in $Li_6C_{60}$ and $Li_6C_{60}H_y$, Carbon, 96 (2016) 276-284. http://dx.doi.org/10.1016/j.carbon.2015.09.064
[39] M. Gaboardi, C. Cavallari, G. Magnani, D. Pontiroli, S. Rols, M. Riccò, Hydrogen storage mechanism and lithium dynamics in Li12C60 investigated by μSR, Carbon, 90 (2015) 130-137. http://dx.doi.org/http://dx.doi.org/10.1016/j.carbon.2015.03.072
[40] D.W.H. Swenson, H.M. Jaeger, C.E. Dykstra, Clustering of molecular hydrogen around benzene, Chem Phys, 326 (2006) 329-334. http://dx.doi.org/http://dx.doi.org/10.1016/j.chemphys.2006.02.009
[41] C. Leidlmair, P. Bartl, H. Schöbel, S. Denifl, M. Probst, P. Scheier, O. Echt, On the Possible Presence of Weakly Bound Fullerene-$H_2$ Complexes in the Interstellar Medium, Astrophys J Lett, 738 (2011) L4.
[42] S. Ralser, J. Postler, M. Harnisch, A.M. Ellis, P. Scheier, Extracting cluster distributions from mass spectra: IsotopeFit, Int J Mass Spectrom, 379 (2015) 194-199. http://dx.doi.org/10.1016/j.ijms.2015.01.004
[43] S. Jaksch, A. Mauracher, A. Bacher, S. Denifl, F. Ferreira da Silva, H. Schöbel, O. Echt, T.D. Märk, M. Probst, D.K. Bohme, P. Scheier, Formation of even-numbered hydrogen cluster cations in ultracold helium droplets, J Chem Phys, 129 (2008) 224306. http://dx.doi.org/http://dx.doi.org/10.1063/1.3035833
[44] A. Mauracher, M. Daxner, J. Postler, S.E. Huber, S. Denifl, P. Scheier, J.P. Toennies, Detection of Negative Charge Carriers in Superfluid Helium Droplets: The Metastable Anions He*$^-$ and He2*$^-$, J Phys Chem Lett, 5 (2014) 2444-2449. http://dx.doi.org/10.1021/jz500917z
[45] F. Ancilotto, E. Cheng, M.W. Cole, F. Toigo, The binding of alkali atoms to the surfaces of liquid helium and hydrogen, Z Physik B - Condensed Matter, 98 (1995) 323-329. http://dx.doi.org/10.1007/BF01338398
[46] M. Theisen, F. Lackner, W.E. Ernst, Cs atoms on helium nanodroplets and the immersion of Cs+ into the nanodroplet, J Chem Phys, 135 (2011) 074306. http://dx.doi.org/doi:http://dx.doi.org/10.1063/1.3624840
[47] E. Cheng, M.W. Cole, W.F. Saam, J. Treiner, Helium prewetting and nonwetting on weak-binding substrates, Phys Rev Lett, 67 (1991) 1007-1010. http://dx.doi.org/https://doi.org/10.1103/PhysRevLett.67.1007
[48] O. Bünermann, M. Mudrich, M. Weidemüller, F. Stienkemeier, Spectroscopy of Cs attached to helium nanodroplets, J Chem Phys, 121 (2004) 8880-8886. http://dx.doi.org/doi:http://dx.doi.org/10.1063/1.1805508
[49] M. Renzler, M. Daxner, L. Kranabetter, A. Kaiser, A.W. Hauser, W.E. Ernst, A. Lindinger, R. Zillich, P. Scheier, A.M. Ellis, Communication: Dopant-induced solvation of alkalis in liquid helium nanodroplets, J Chem Phys, 145 (2016) 181101. http://dx.doi.org/doi:http://dx.doi.org/10.1063/1.4967405
[50] A.A. Scheidemann, V.V. Kresin, H. Hess, Capture of lithium by 4He clusters: Surface adsorption, Penning ionization, and formation of HeLi+, J Chem Phys, 107 (1997) 2839-2844. http://dx.doi.org/doi:http://dx.doi.org/10.1063/1.474642
[51] S.E. Huber, A. Mauracher, On the properties of charged and neutral, atomic and molecular helium species in helium nanodroplets: interpreting recent experiments, Mol Phys, 112 (2014) 794-804. http://dx.doi.org/10.1080/00268976.2013.863403

[52] F. Rabilloud, Structure and Electronic Properties of Alkali-$C_{60}$ Nanoclusters, J Phys Chem A, 114 (2010) 7241-7247. http://dx.doi.org/10.1021/jp103124w

[53] M. Robledo, F. Martin, M. Alcami, S. Diaz-Tendero, Exohedral interaction in cationic lithium metallofullerenes, Theor Chem Acc, 132 (2013) 1346. http://dx.doi.org/10.1007/s00214-013-1346-8

[54] B.K. Rao, P. Jena, Hydrogen Uptake by an Alkali Metal Ion, Europhysics Letters, 20 (1992) 307. http://dx.doi.org/http://dx.doi.org/10.1209/0295-5075/20/4/004

[55] A. Kaiser, C. Leidlmair, P. Bartl, S. Zöttl, S. Denifl, A. Mauracher, M. Probst, P. Scheier, O. Echt, Adsorption of hydrogen on neutral and charged fullerene: Experiment and theory, J Chem Phys, 138 (2013) 074311-074313. http://dx.doi.org/10.1063/1.4790403

[56] L.G. Scanlon, W.A. Feld, P.B. Balbuena, G. Sandi, X. Duan, K.A. Underwood, N. Hunter, J. Mack, M.A. Rottmayer, M. Tsao, Hydrogen Storage Based on Physisorption, J Phys Chem B, 113 (2009) 4708-4717. http://dx.doi.org/10.1021/jp809097v

[57] F. Calvo, Size-induced melting and reentrant freezing in fullerene-doped helium clusters, Phys Rev B, 85 (2012) 060502.

[58] S. Zöttl, A. Kaiser, M. Daxner, M. Goulart, A. Mauracher, M. Probst, F. Hagelberg, S. Denifl, P. Scheier, O. Echt, Ordered phases of ethylene adsorbed on charged fullerenes and their aggregates, Carbon, 69 (2014) 206-220. http://dx.doi.org/http://dx.doi.org/10.1016/j.carbon.2013.12.017

[59] A. Kaiser, S. Zöttl, P. Bartl, C. Leidlmair, A. Mauracher, M. Probst, S. Denifl, O. Echt, P. Scheier, Methane Adsorption on Aggregates of Fullerenes: Site-Selective Storage Capacities and Adsorption Energies, ChemSusChem, 6 (2013) 1235 – 1244. http://dx.doi.org/10.1002/cssc.201300133

[60] F. Calvo, E. Yurtsever, Coating of $C_{60}$ by para-$H_2$ and ortho-$D_2$ : revisiting the solvation shell—CMMSE, Journal of Mathematical Chemistry, (2016) 1-6. http://dx.doi.org/10.1007/s10910-016-0679-7

[61] L.F. Gomez, E. Loginov, R. Sliter, A.F. Vilesov, Sizes of large He droplets, J Chem Phys, 135 (2011) 154201. http://dx.doi.org/doi:http://dx.doi.org/10.1063/1.3650235

[62] J.P. Toennies, A.F. Vilesov, Superfluid Helium Droplets: A Uniquely Cold Nanomatrix for Molecules and Molecular Complexes, Angew Chemie (Int Ed), 43 (2004) 2622-2648.

[63] H. Schöbel, P. Bartl, C. Leidlmair, S. Denifl, O. Echt, T.D. Märk, P. Scheier, High-Resolution Mass Spectrometric Study of Pure Helium Droplets, and Droplets Doped with Krypton, Eur Phys J D, 63 (2011) 209-214. http://dx.doi.org/DOI: 10.1140/epjd/e2011-10619-1

[64] S. Ralser, J. Postler, M. Harnisch, A.M. Ellis, P. Scheier, Extracting Cluster Distributions from Mass Spectra: IsotopeFit, Int J Mass Spectrom, 379 (2015) 194-199. http://dx.doi.org/doi:10.1016/j.ijms.2015.01.004

[65] J.-D. Chai, M. Head-Gordon, Long-range corrected hybrid density functionals with damped atom-atom dispersion corrections, Phys Chem Chem Phys, 10 (2008) 6615-6620. http://dx.doi.org/http://dx.doi.org/10.1039/b810189b

[66] S. Grimme, Semiempirical GGA-type density functional constructed with a long-range dispersion correction, J Comp Chem, 27 (2006) 1787-1799. http://dx.doi.org/http://dx.doi.org/10.1002/jcc.20495

[67] C. Adamo, V. Barone, Toward reliable density functional methods without adjustable parameters: The PBE0 model, J Chem Phys, 110 (1999) 6158-6170. http://dx.doi.org/doi:http://dx.doi.org/10.1063/1.478522

[68] J.P. Perdew, K. Burke, M. Ernzerhof, Generalized Gradient Approximation Made Simple, Phys Rev Lett, 77 (1996) 3865-3868. http://dx.doi.org/https://doi.org/10.1103/PhysRevLett.77.3865

[69] S. Grimme, J. Antony, S. Ehrlich, H. Krieg, A consistent and accurate ab initio parametrization of density functional dispersion correction (DFT-D) for the 94 elements H-Pu, J Chem Phys, 132 (2010) 154104. http://dx.doi.org/doi:http://dx.doi.org/10.1063/1.3382344

[70] S. Grimme, S. Ehrlich, L. Goerigk, Effect of the damping function in dispersion corrected density functional theory, J Comput Chem, 32 (2011) 1456-1465. http://dx.doi.org/10.1002/jcc.21759

[71] T.H. Dunning Jr., P.J. Hay, Modern Theoretical Chemistry, New York, Plenum, 1977.

[72] W.R. Wadt, P.J. Hay, Ab initio effective core potentials for molecular calculations. Potentials for main group elements Na to Bi, J Chem Phys, 82 (1985) 284-298. http://dx.doi.org/doi:http://dx.doi.org/10.1063/1.448800

[73] K.L. Schuchardt, B.T. Didier, T. Elsethagen, L. Sun, V. Gurumoorthi, J. Chase, J. Li, T.L. Windus, Basis Set Exchange: A Community Database for Computational Sciences, J Chem Inf Model, 47 (2007) 1045-1052. http://dx.doi.org/10.1021/ci600510j

[74] M.J. Frisch, G.W. Trucks, H.B. Schlegel, G.E. Scuseria, M.A. Robb, J.R. Cheeseman, G. Scalmani, V. Barone, B. Mennucci, G.A. Petersson, H. Nakatsuji, M. Caricato, X. Li, H.P. Hratchian, A.F. Izmaylov, J. Bloino, G. Zheng, J.L. Sonnenberg, M. Hada, M. Ehara, K. Toyota, R. Fukuda, J. Hasegawa, M.I.a. T., Nakajima, Y. Honda, O. Kitao, H. Nakai, T. Vreven, J.A. Montgomery, J.E. Peralta, F. Ogliaro, M. Bearpark, J.J. Heyd, E. Brothers, K.N. Kudin, V.N. Staroverov, R. Kobayashi, J. Normand, K. Raghavachari, A. Rendell, J.C. Burant, S.S. Iyengar, J. Tomasi, M. Cossi, N. Rega, J.M. Millam, M. Klene, J.E. Knox, J.B. Cross, V. Bakken, C. Adamo, J. Jaramillo, R. Gomperts, R.E. Stratmann, O. Yazyev, A.J. Austin, R. Cammi, C. Pomelli, J.W. Ochterski, R.L. Martin, K. Morokuma, V.G.

Zakrzewski, G.A. Voth, P. Salvador, J.J. Dannenberg, S. Dapprich, A.D. Daniels, Ö. Farkas, J.B. Foresman, J.V. Ortiz, J. Cioslowski, D.J. Fox, Gaussian 09 Revision D.01, in, Gaussian Inc. Wallingford CT 2009.